\documentclass[11pt,twoside]{article}

\usepackage{asp2006}
\usepackage{epsf}
\usepackage{psfig}
\usepackage{graphicx}
\usepackage{lscape}
\usepackage{amsbsy}
\usepackage{amssymb}

\begin{document}
%%% Fill in title
\title{Observational signatures of  numerically simulated MHD
waves in small-scale fluxtubes}

%%% Fill in author names
\author{E. Khomenko\altaffilmark{1,2}, M. Collados\altaffilmark{1} and
        T. Felipe\altaffilmark{1}}

%%% Fill in author affiliations
\affil{(1) Instituto de Astrof\'{\i}sica de Canarias, 38205, C/
V\'{\i}a L{\'a}ctea, s/n, Tenerife, Spain}

\affil{(2) Main Astronomical Observatory, NAS, 03680 Kyiv,
Zabolotnogo str. 27, Ukraine}

%%% Abstract to run on from here.
\begin{abstract}
We present some results obtained from the synthesis of Stokes
profiles in small-scale flux tubes with propagating MHD waves. To
that aim, realistic flux tubes showing internal structure have
been excited with 5 min period drivers, allowing non-linear waves
to propagate inside the magnetic structure. The observational
signatures of these waves in Stokes profiles of several spectral
lines that are commonly used in spectropolarimetric measurements
are discussed.
\end{abstract}

\section{Introduction}

The observed wave behaviour in network and facular solar regions,
and the role of waves to connect photospheric and chromospheric
layers, have drawn a considerable discussion in the recent
literature, since it has been proposed that thin magnetic flux
tubes in these regions can act as wave guides, supplying energy to
the upper layers, in particular for waves with periodicity in the
5-min range \citep{DePontieu+etal2004}. However, observational
studies of the relation between the photospheric and chromospheric
signals do not provide a unique picture. It is not clear whether
the oscillations remain coherent through the atmosphere
\citep{Lites+Rutten+Kalkofen1993, Krijer+etal2001, Judge+etal2001,
Centeno+etal2006b}. There are some hints that the coupling between
the photosphere and the chromosphere is within vertical channels
\citep{Judge+etal2001, Centeno+etal2006b}, however, other results
point toward inclined wave propagation \citep{Bloomfield+etal2006,
DePontieu+etal2003}. In particular, all these studies point out
the important role of the magnetic canopy and the $c_S=v_A$
transformation layer. The dominance of long-period oscillations in
the chromosphere of these regions, also remains to be explained.

Here we perform numerical simulations of magneto-acoustic waves in
small-scale flux tubes, with a subsequent Stokes diagnostics. We
address the questions of coherence of oscillations through the
atmosphere, wave behaviour at the canopies and the change of the
wave period with height.

\section{Simulations and spectral synthesis}

We have solved numerically the set of non-linear 2D MHD equations
using the code described in \citet{Khomenko+Collados2006,
Khomenko+Collados2007}.
As initial condition, we have used a magnetostatic flux tube model
constructed following the method by \citet{Pneuman+etal1986}, with
a radius of 100 km at the base of the photosphere. This model is a
second order approximation to a thin flux tube and has both
vertical and horizontal variations in all atmospheric parameters.
The magnetic field along the flux tube axis varies with height
from 740 G in the photosphere to 37 G at z=2000 km, and the plasma
$\beta$ decreases from 4 to 7$\times 10^{-4}$.
We compare here three simulation runs. In the first one,
oscillations were excited by a photospheric driver with horizontal
velocity given by $V_x(x,\it{t})=V_0 \sin(\omega \it{t})\times
\exp(-(x-x_0)^2/2\sigma^2)$, where $\sigma=160$ km, $2
\pi/\omega=300$ sec, $V_0=200$ \hbox{m$\;$s$^{-1}$} is the initial
velocity amplitude at the photosphere and $x_0$ corresponds to the
position of the tube axis. This run was performed in an adiabatic
regime.
In the other two runs, oscillations were excited by a periodic
photospheric driver in vertical velocity with the same horizontal
and temporal dependence as before. One of these runs was adiabatic
while the other was carried out with a radiative relaxation time
of $\tau_{RR}=10$ s (constant through the whole atmosphere). The
radiative losses were described by Newton's law of cooling.
The Stokes spectra of the \mbox{Fe\,\sc{i}} pair of lines at 1.565
$\mu$m and of the \mbox{Si\,\sc{i}} line at 1.083 $\mu$m were
synthesized with the help of the SIR code
\citep{RuizCobo+delToroIniesta1992} for each horizontal position
and time step as emerging at solar disk center.
%According to their formation heights, the \FeI\ lines give
%information about the deep photosphere and the \SiI\ line about
%the high photosphere close to the temperature minimum.

%%%%%%%%%%%%%%%%%%%%%%%%%%%%%%%%%%%%%%%%%%%%%%%%%%%%%%%%%%%%fig1
\begin{figure*}[!t]
\centering
 \includegraphics[width=0.6\textwidth]{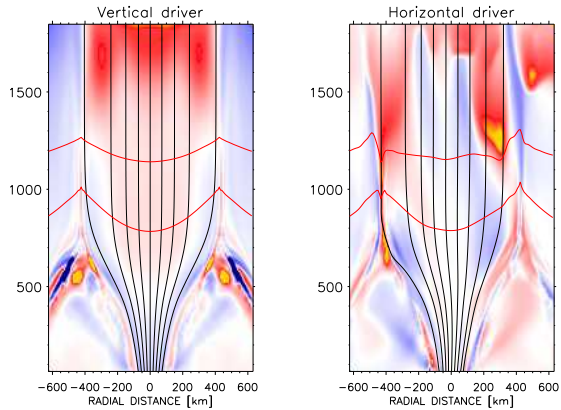}
\caption{Snapshots of the vertical velocity (normalized to
$\rho_0^{1/4}$) in adiabatic simulations excited with a vertical
driver (left) and a horizontal driver (right) after about 800 sec
elapsed time.  The vertical lines represent the magnetic field
lines. The two horizontal lines are contours of constant
$c_S^2/v_A^2$, the thick line corresponding to $v_A=c_S$ and the
thin line to $c_S^2/v_A^2=0.1$. } \label{fig:snap}
\end{figure*}
%%%%%%%%%%%%%%%%%%%%%%%%%%%%%%%%%%%%%%%%%%%%%%%%%%%%%%%%%%%%fig1

Fig.~\ref{fig:snap} compares two snapshots of the adiabatic
simulations with a vertical (left) and horizontal (right) driver.
The vertical driver excites in the photosphere the fast (acoustic)
mode and the surface mode. The fast mode is transformed into the
slow acoustic mode after the $c_S=v_A$ height and continues up to
the chromosphere forming shocks with an amplitude of 3--5
\hbox{km$\;$s$^{-1}$}. Note that, in the case of the vertical
driver, all variations are approximately in phase on both sides of
the flux tube. The horizontal driver excites in the photosphere
the slow (magnetic) mode and also the surface mode. The slow mode
produces shaking of the tube in horizontal direction. After the
transformation layer, most of the energy of the slow magnetic mode
goes, again, to the slow acoustic mode. Note that, in these
simulations with the horizontal driver, the variations are
approximately in antiphase on both sides of the flux tube at
photospheric heights. For more details on these simulations, see
\citet{Khomenko+Collados2007}.

\section{Surface mode and acoustic mode signatures}

%%%%%%%%%%%%%%%%%%%%%%%%%%%%%%%%%%%%%%%%%%%%%%%%%%%%%%%%%%%%fig1
\begin{figure*}[!t]
\plottwo{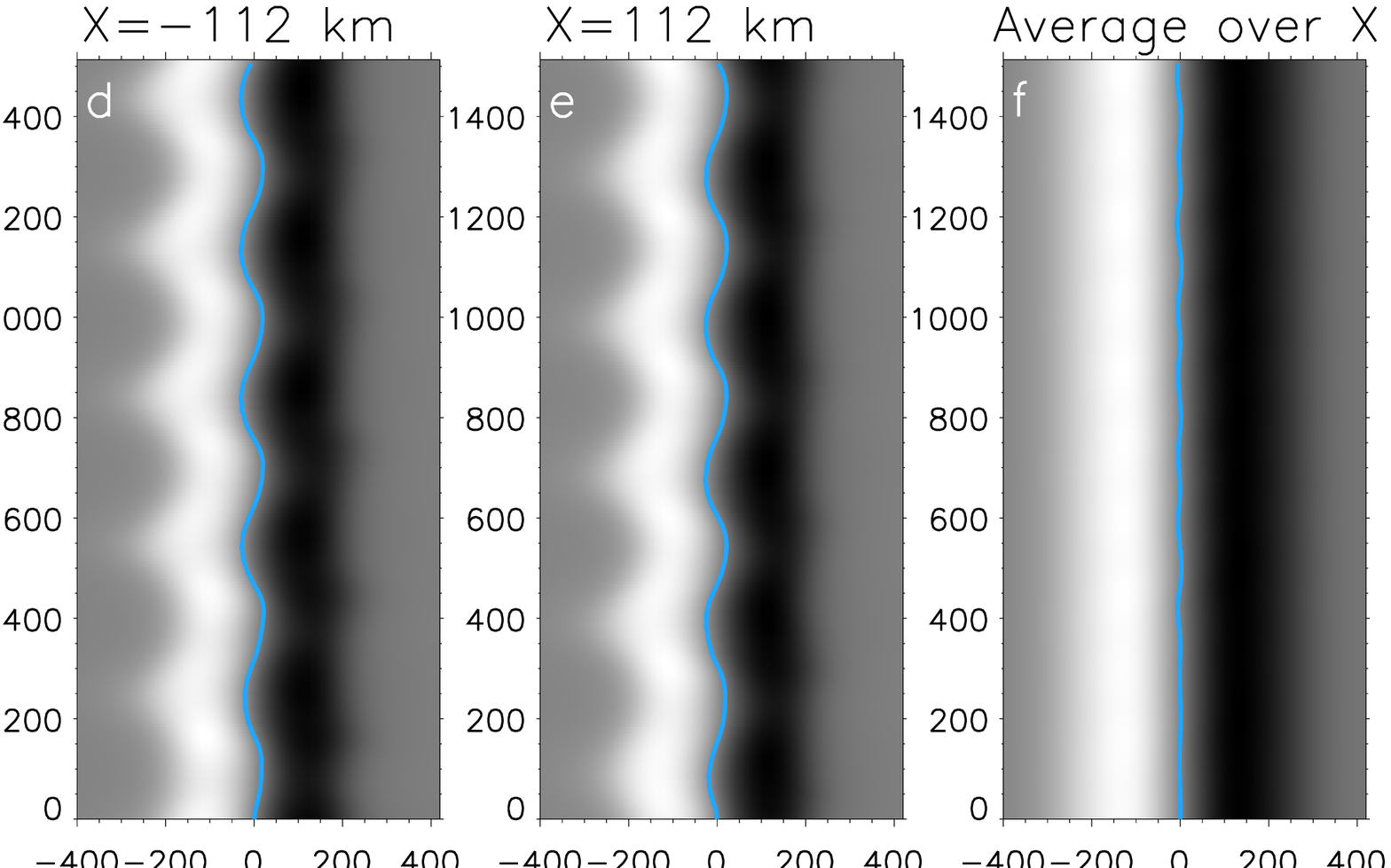}{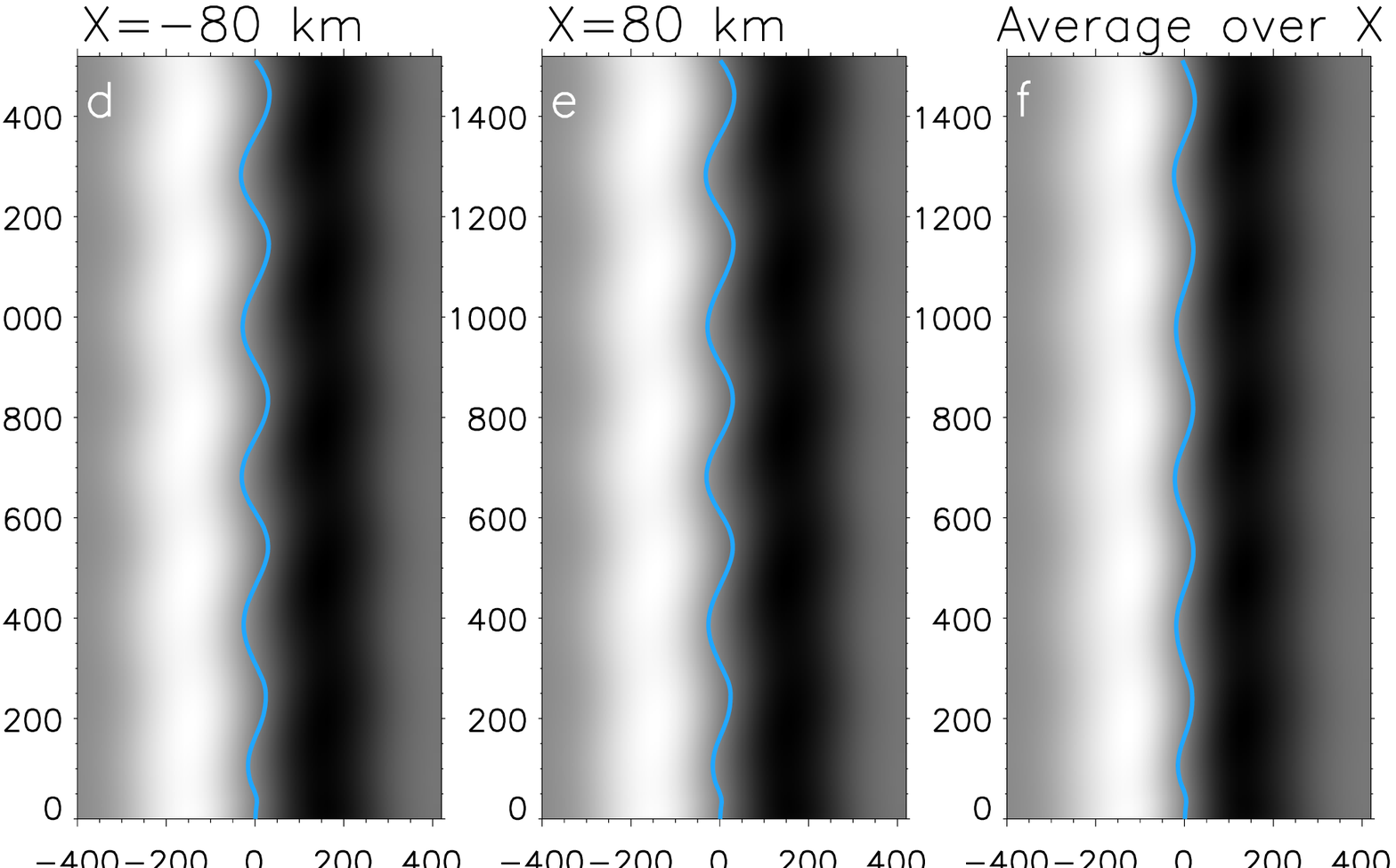}
\caption{Time evolution of Stokes $V$ profiles of the
\mbox{Fe\,\sc{i}} 1564.8 nm line in the adiabatic simulations with
a horizontal (panels a, b, and c) and a vertical (panels d, e and
f) driver. In each group, the two first panels give spectra at two
symmetric positions with respect to the tube axis, while the third
one gives the spatially averaged spectra over the simulation
domain. } \label{fig:time_wl}
\end{figure*}
%%%%%%%%%%%%%%%%%%%%%%%%%%%%%%%%%%%%%%%%%%%%%%%%%%%%%%%%%%%%fig1

Fig.~\ref{fig:time_wl} gives the time evolution of the Stokes $V$
spectra of the \mbox{Fe\,\sc{i}} 1564.8 nm line for different
cases (see figure caption). In the case of the simulations with a
horizontal driver the variations of the synthesized Stokes
profiles are due to the surface wave. The slow mode is transversal
and does not produce any significant variations of the vertical
velocity. The variations of amplitude and zero-crossing shift of
the Stokes $V$ profiles are in antiphase. As can be observed from
panel (c) of Fig.~\ref{fig:time_wl}, these variations cancel out
in the case of observations with insufficient spatial resolution.
Thus, no time variations of the velocity (nor of any other
parameter characteristic of the profiles) would be detected in the
photosphere at the solar disc center if the driver that produces
oscillations is purely horizontal.

In the case of the simulations with the vertical driver, time
variations of the Stokes $V$ profiles are mainly caused by the
fast acoustic mode. The oscillations are in phase on both sides of
the tube. Thus, spatial averaging does not affect the resulting
variations to a large extent (Fig.~\ref{fig:time_wl}f). An
observer would detect low spatial resolution velocity oscillations
with an amplitude of some 150 \hbox{m$\;$s$^{-1}$} and Stokes $V$
amplitude variations of 10$^{-3}$, in units of the continuum
intensity. Note also that, according to Fig.~\ref{fig:snap} (left
panel), the phase of the oscillations changes rather slowly with
height. Thus, oscillations detected in the photosphere and the
chromosphere would be coherent (with some delay) if the driver
that excites oscillations has a vertical component.

%%%%%%%%%%%%%%%%%%%%%%%%%%%%%%%%%%%%%%%%%%%%%%%%%%%%%%%%%%%%fig1
\begin{figure*}[!t]
 \plotone{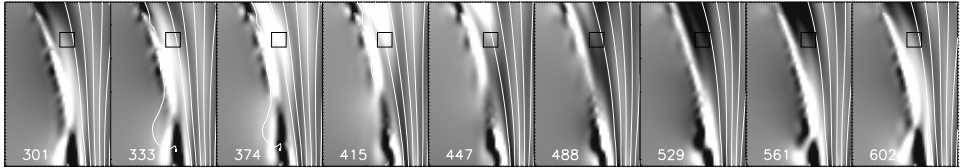} \caption{Time evolution of the vertical velocity, in
 a part of the domain (300$\times$400 km), from the simulations with a horizontal driver, showing the
 wave traveling at the canopy zone.  Numbers at each panel give the elapsed time in seconds. The small
 square refers to a position where velocities are similar to those derived from
 the V-profiles of the \mbox{Si\,\sc{i}} line (see Fig. 4).}
\label{fig:canopy_time}
\end{figure*}
%%%%%%%%%%%%%%%%%%%%%%%%%%%%%%%%%%%%%%%%%%%%%%%%%%%%%%%%%%%%fig1

\section{Waves at canopies}

\begin{figure*}[!t]
\centering
\includegraphics[width=0.6\textwidth]{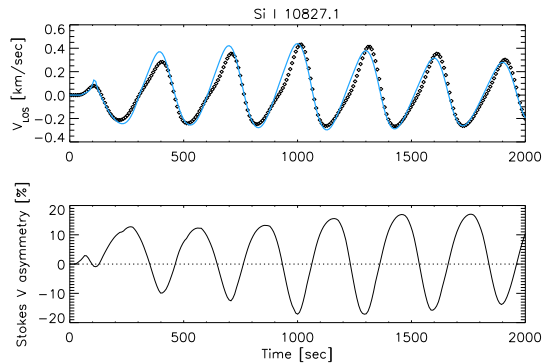}
\caption{Top: diamonds give time variation of the zero-crossing
velocity of the \mbox{Si\,\sc{i}} Stokes $V$ profiles (positive
velocities meaning downward motion). The solid curve gives the
vertical velocity extracted directly from the simulations at
position marked by black square in Fig.~\ref{fig:canopy_time}.
Bottom: time evolution of the Stokes $V$ amplitude asymmetry.}
\label{fig:can}
\end{figure*}

Horizontal photospheric shaking of the flux tube due to the slow
mode in the simulations with a horizontal driver produces strong
compression on the interfaces between the flux tube and its
non-magnetic surroundings. Due to that, already at the heights of
formation of the \mbox{Si\,\sc{i}} line, the vertical velocity
variations induced by the surface mode show a weak non-linear
behaviour, which is illustrated in Figs.~\ref{fig:canopy_time} and
\ref{fig:can}.
%The former figure gives the time evolution of a
%part of the simulation domain containing the region of interest.
%Some parameters calculated from these profiles are plotted in
%Fig.~\ref{fig:can}.

%%%%%%%%%%%%%%%%%%%%%%%%%%%%%%%%%%%%%%%%%%%%%%%%%%%%%%%%%%%%fig1
%%%%%%%%%%%%%%%%%%%%%%%%%%%%%%%%%%%%%%%%%%%%%%%%%%%%%%%%%%%%fig1

An inspection of the temporal evolution of the simulations
(Figs.~\ref{fig:canopy_time}) reveals an increase of the vertical
velocity amplitude just at the border between the flux tube and
the field free atmosphere and the absence of any significant
velocity variations outside the flux tube. These abrupt changes
produce a weak shock wave behaviour of the zero-crossing velocity
calculated from the Stokes $V$ profiles of the \mbox{Si\,\sc{i}}
line (upper panel of Fig.~\ref{fig:can}). For comparison, we also
show in this figure a velocity curve extracted directly from the
simulations at the corresponding height. The similarity between
both curves is evident, confirming that the variations of the
\mbox{Si\,\sc{i}} Stokes $V$ zero-crossing position are related to
heights about 300-350 km in the photosphere. The magnetic field
discontinuity, together with the velocity variations, at the
field-free interface produce strong asymmetries of the Stokes $V$
profiles. The variation of the Stokes $V$ amplitude asymmetry can
be as large as $\pm$20\% (bottom panel of Fig.~\ref{fig:can}). As
expected, the asymmetry is anticorrelated with the zero-crossing
velocity, being maximum when the velocity reaches maximum upward
values.

%%%%%%%%%%%%%%%%%%%%%%%%%%%%%%%%%%%%%%%%%%%%%%%%%%%%%%%%%%%%fig1
\begin{figure*}[!t]
\centering
 \includegraphics[width=0.6\textwidth]{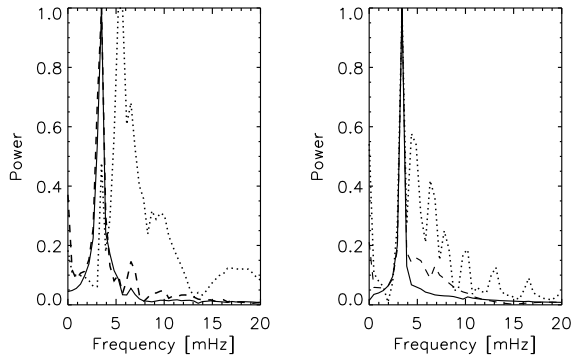}
 \caption{Normalized power spectra of the LOS velocity oscillations in
 the \mbox{Fe\,\sc{i}} 1564.8 nm line (solid line), in the \mbox{Si\,\sc{i}} line (dashed line) and in
 the simulations at a height of about 1500 km (dotted line), averaged over the central
 part of the tube. Left panel: adiabatic simulations with a vertical driver;
 right panel: simulations with the same driver,
 but including radiative losses.} \label{fig:spectra}
\end{figure*}
%%%%%%%%%%%%%%%%%%%%%%%%%%%%%%%%%%%%%%%%%%%%%%%%%%%%%%%%%%%%fig1

\section{Power spectra at different heights}

In the above sections, adiabatic simulations have been discussed.
The simulations with a vertical driver including radiative losses
are qualitatively similar to those without radiative losses. In
the photosphere, both the surface and the fast modes are generated
and fast-to-slow mode transformation is observed at the $c_S=v_A$
layer. However, there is an important difference. In the adiabatic
simulations, the frequency of the oscillations changes with
height, due to the non-linear generation of the secondary
harmonics, from 3 mHz to 6 mHz. The oscillations excited by the
driver at 3 mHz are evanescent already in the photosphere and,
thus, can not propagate, unlike their higher frequency harmonics.
This is not the situation if oscillations are allowed to exchange
radiative energy with their surroundings. Including energy losses
is formally equal to reducing the effective cut-off frequency of
the atmosphere by a certain amount \citep{Roberts1983,
Centeno+etal2006b, Khomenko+etal2008}. Thus, waves at 3 mHz can
propagate through the whole atmosphere.

Fig.~\ref{fig:spectra} gives the power spectra of oscillations at
different heights. The solid curve is obtained from the
\mbox{Fe\,\sc{i}} line profiles, i.e. corresponds to the deep
photosphere. There, the dominating frequency is that of the
driver, 3 mHz. Higher up, in both, adiabatic and non-adiabatic
vertically-driven, simulations, power at higher frequencies starts
to appear. This can be appreciated from the dashed curve obtained
from the \mbox{Si\,\sc{i}} profiles. There is still not much
difference between the adiabatic and non-adiabatic simulations at
photospheric heights. The dotted curves correspond to the vertical
velocity variations at a height of about 1500 km, in the
chromosphere. Now, the adiabatic simulations show a clear shift of
the power to higher frequencies, while non-adiabatic simulations
have the dominant oscillations at the frequency of the driver at 3
mHz.
This implies that vertical thin magnetic structures, such as flux
tubes, can channel 5-minute oscillations from the photosphere to
the chromosphere if radiative effects are taken into account.

\section{Conclusions}

In this paper we have studied the wave behaviour in small-scale
flux tubes by means of non-linear numerical simulations and Stokes
diagnostics. The following items summarize our conclusions.

(i) If the driver that excites oscillations has a vertical
component, velocity oscillations with an amplitude of some 150
\hbox{m$\;$s$^{-1}$} and Stokes $V$ amplitude oscillations of
10$^{-3}$, in units of the continuum intensity,  may be detected
in low spatial resolution observations. The oscillations are
coherent through the whole atmosphere (with the corresponding time
delay due the upward propagation).

(ii) If the driver that excites oscillations is purely horizontal,
no variations would be detected in observations with reduced
spatial resolution, since the vertical velocity variations are
produced by the surface mode and are in antiphase on both sides of
the tube.

(iii) The LOS velocity shows a weak non-linear behaviour in canopy
regions, already at heights of 300--350 km in the photosphere.

(iv) It is important to take into account the radiative losses of
oscillations. They can produce a decrease of the effective cut-off
frequency, making possible the propagation of the 5-min
oscillations in the chromosphere in vertical small-scale magnetic
structures.

%%% Text of acknowledgements runs on after this command.
\acknowledgements{Financial support by the European Commission
through the SOLAIRE Network (MTRN-CT-2006-035484) and by the
Spanish Ministery of Education through projects AYA2007-66502 and
AYA2007-63881 is gratefully acknowledged.}

%%% THE BIBLIOGRAPHY
%%%
%%% CONSULT SECTION 3 OF "INSTRUCTIONS FOR AUTHORS" FOR HOW TO USE NATBIB.
%%% AUTHORS ARE ENCOURAGED TO USE EITHER THE "THEBIBLIOGRAPY" ENVIRONMENT
%%% BY UNCOMMENTING (DELETING THE "%" SYMBOL) THE COMMANDS BELOW, OR BY
%%% USING THE BIBTEX ENVIRONMENT. TO FIND OUT WHICH IS APPLICABLE TO YOUR
%%% CONTRIBUTION, CONSULT THE VOLUME EDITORS FOR YOUR PROCEEDINGS.
%%%


\begin{thebibliography}{}

\bibitem[\protect\astroncite{Bloomfield et~al.}{2006}]{Bloomfield+etal2006}
Bloomfield, D.~S., McAteer, R. T.~J., Mathioudakis, M., Keenan, F.~P. 2006,
  ApJ, 652, 812

\bibitem[\protect\astroncite{Centeno et~al.}{2006}]{Centeno+etal2006b}
Centeno, R., Collados, M., \mbox{Trujillo Bueno}, J. 2006,
\newblock in R. Casini, B.~W. Lites (eds.), Solar Polarization 4, Vol. 358, ASP
  Conference Series,  465

\bibitem[\protect\astroncite{Judge et~al.}{2001}]{Judge+etal2001}
Judge, P.~G., Tarbell, T.~D., Wilhelm, K. 2001, ApJ, 554, 424

\bibitem[\protect\astroncite{Khomenko et~al.}{2008}]{Khomenko+etal2008}
Khomenko, E., Centeno, R., Collados, M., \mbox{Trujillo Bueno}, J. 2008, ApJ,
  submitted

\bibitem[\protect\astroncite{Khomenko \&
  Collados}{2006}]{Khomenko+Collados2006}
Khomenko, E., Collados, M. 2006, ApJ, 653, 739

\bibitem[\protect\astroncite{Khomenko \&
  Collados}{2007}]{Khomenko+Collados2007}
Khomenko, E., Collados, M. 2007, Solar Phys., submitted

\bibitem[\protect\astroncite{Krijger et~al.}{2001}]{Krijer+etal2001}
Krijger, J.~M., Rutten, R.~J., Lites, B.~W., Straus, T., Shine, R.~A., Tarbell,
  T.~D. 2001, A\&A, 379, 1052

\bibitem[\protect\astroncite{Lites et~al.}{1993}]{Lites+Rutten+Kalkofen1993}
Lites, B.~W., Rutten, R.~J., Kalkofen, W. 1993, ApJ, 414, 345

\bibitem[\protect\astroncite{\mbox{De Pontieu}
  et~al.}{2003}]{DePontieu+etal2003}
\mbox{De Pontieu}, B., Erdelyi, R., de~Wijn, A.~G. 2003, ApJ, 595, L63

\bibitem[\protect\astroncite{\mbox{De Pontieu}
  et~al.}{2004}]{DePontieu+etal2004}
\mbox{De Pontieu}, B., Erdelyi, R.~J., Stewart, P. 2004, Nat, 430, Issue 6999,
  536

\bibitem[\protect\astroncite{\mbox{Ruiz Cobo} \& \mbox{del Toro
  Iniesta}}{1992}]{RuizCobo+delToroIniesta1992}
\mbox{Ruiz Cobo}, B., \mbox{del Toro Iniesta}, J.~C. 1992, ApJ, 398, 375

\bibitem[\protect\astroncite{Pneuman et~al.}{1986}]{Pneuman+etal1986}
Pneuman, G.~W., Solanki, S.~K., Stenflo, J.~O. 1986, A\&A, 154, 231

\bibitem[\protect\astroncite{Roberts}{1983}]{Roberts1983}
Roberts, B. 1983, Solar Phys., 87, 77

\end{thebibliography}
\end{document}